\documentclass{ws-procs9x6}

\begin{document}

\title{Exact Standard Model Structures from
Intersecting Branes}

\author{C.~KOKORELIS}

\address{ Departamento de F\'\i sica Te\'orica C-XI and 
Instituto de F\'\i sica 
Te\'orica C-XVI\\
Universidad Aut\'onoma de Madrid, Cantoblanco, 28049, Madrid, Spain\\ 
E-mail: C.Kokorelis@uam.es}


\maketitle

\abstracts{I discuss two types of non-supersymmetric string model
constructions
that give at low energy
exactly the Standard model (SM) with no additional matter/and or gauge group 
factors. The construction is based on D6 branes intersecting
at angles in a compactification of type IIA theory on a
decomposable orientifolded $T^6$ torus. 
The first type 
is based on five and six stack SM-like constructions at the
string scale while, the other construction is based on
a four stack GUT left-right symmetric structure centered around the
Pati-Salam $SU(4)_C \times SU(2)_L \times SU(2)_R$ gauge group.
All classes of models exbibit important
phenomenological properties including a stable proton and
sizes of neutrino masses in consistency with neutrino oscillation
experiments. The models are non-SUSY, but amazingly, they allow the existence
of supersymmetric particles!}

\section{Introduction}
One of most difficult tasks that string theory has to phase 
today \footnote{In the absence of a dynamical mechanism
that can select a particular string vacuum.} is the
construction of non-SUSY vacua with exactly the observable SM gauge group and 
interactions at low energies of order $M_z$. 
The latter had become possible only recently,
 as vacua with exactly the SM at low 
energy have been constructed either from D6 branes intersecting
over \cite{tessera,tessera2} an orientifolded $T^6$ 
torus \cite{mo1,mo2,mo22,mo3,mo4} or from intersecting D5 
branes on an orientifold of $T^2 \times T^4/Z_N$ \cite{}\cite{mo5,mo6}.
For constructions of SUSY vacua\footnote{In the context of intersecting 
branes.} see \cite{mo7}.  
In this talk, we will focused on the four stack GUT constructions 
of \cite{mo2} as well the
the five and six stack SM 
constructions of \cite{mo3,mo4}.

\section{Five Stack SM's }
The models start with five stacks of D6 branes making a
 $U(3) \times U(2) \times U(1)_c \times U(1)_d
\times U(1)_e$ group structure at the string scale 
\footnote{Effectively an $SU(3) \times SU(2) \times U(1)_a \times U(1)_b 
\times U(1)_c \times U(1)_d \times U(1)_e$, as
each $U(N_i)$ will give rise to an $SU(N_i)$
charged under the associated $U(1_i)$ gauge group factor that appears in 
the decomposition $SU(N_a) \times U(1_a)$.}
The models are constructed as a deformation of the four stack models
of \cite{mo1} around the QCD intersection numbers $I_{ab^{\star}}= 2$, 
$I_{ab}= 1$, $I_{ac^{\star}}= -3$, $I_{ac}= 3$.
The SM spectrum is localized in the intersections as in 
table (\ref{spectrum8}).
\begin{table}[th] 
\tbl{
Low energy fermionic spectrum of the five stack 
string scale 
$SU(3)_C \otimes
SU(2)_L \otimes U(1)_a \otimes U(1)_b \otimes U(1)_c 
\otimes U(1)_d \otimes U(1)_e $, type I D6-brane model together with its
$U(1)$ charges. Note that at low energies only the SM gauge group 
$SU(3) \otimes SU(2)_L \otimes U(1)_Y$ survives.
\vspace*{1pt}}
{\footnotesize
\begin{tabular}{|c|c|c|c|c|c|c|c|c|}
\hline
Matter Fields & & Intersection & $Q_a$ & $Q_b$ & $Q_c$ & $Q_d$ & $Q_e$& Y
\\\hline
 $Q_L$ &  $(3, 2)$ & $I_{ab}=1$ & $1$ & $-1$ & $0$ & $0$ & $0$& $1/6$ \\\hline
 $q_L$  &  $2(3, 2)$ & $I_{a b^{\ast}}=2$ &  
$1$ & $1$ & $0$ & $0$  & $0$ & $1/6$  \\\hline
 $U_R$ & $3({\bar 3}, 1)$ & $I_{ac} = -3$ & 
$-1$ & $0$ & $1$ & $0$ & $0$ & $-2/3$ \\\hline    
 $D_R$ &   $3({\bar 3}, 1)$  &  $I_{a c^{\ast}} = -3$ &  
$-1$ & $0$ & $-1$ & $0$ & $0$ & $1/3$ \\\hline    
$L$ &   $2(1, 2)$  &  $I_{bd} = -2$ &  
$0$ & $-1$ & $0$ & $1$ & $0$ & $-1/2$  \\\hline    
$l_L$ &   $(1, 2)$  &  $I_{b e} = -1$ &  
$0$ & $-1$ & $0$ & $0$ & $1$ & $-1/2$  \\\hline    
$N_R$ &   $2(1, 1)$  &  $I_{cd} = 2$ &  
$0$ & $0$ & $1$ & $-1$ & $0$ & $0$  \\\hline    
$E_R$ &   $2(1, 1)$  &  $I_{c d^{\ast}} = -2$ &  
$0$ & $0$ & $-1$ & $-1$ & $0$ & $1$   \\\hline
  $\nu_R$ &   $(1, 1)$  &  $I_{c e} = 1$ &  
$0$ & $0$ & $1$ & $0$ & $-1$ & $0$ \\\hline
$e_R$ &   $(1, 1)$  &  $I_{c e^{\ast}} = -1$ &  
$0$ & $0$ & $-1$ & $0$ & $-1$  & $1$ \\\hline    
\end{tabular}
\label{spectrum8}}
\end{table}
The solution to the RR tadpole cancellation conditions 
\begin{eqnarray}
\sum_a N_a n_a^1 n_a^2 n_a^3 =\ 16,\nonumber\\
\sum_a N_a m_a^1 m_a^2 n_a^3 =\ 0,\nonumber\\
\sum_a N_a m_a^1 n_a^2 m_a^3 =\ 0,\nonumber\\
\sum_a N_a n_a^1 m_a^2 m_a^3 =\ 0.
\label{eq:sp}
\end{eqnarray}

that guarantee the 
absence of non-abelian gauge anomalies
is given in table 1.a.
The choise of tadpole solutions of table 1.a
satisfy all tadpole equations in (\ref{eq:sp}) but the first, 
the latter giving 
\begin{equation}
\frac{9 n_a^2}{ \beta^1} +\ 2 \frac{n_b^1}{ \beta^2} +\
\frac{n_d^2}{ \beta^1} +\ \frac{n_e^2}{ \beta^1} +\
N_D \frac{2}{\beta^1 \beta^2} =\ 16.
\label{eq:ena11}
\end{equation}
The mixed anomalies of the $U(1)$'s with the non-abelian gauge groups
are cancelled through a generalized Green-Schwarz mechanism \cite{mo1} 
that makes massive the $U(1)$'s coupled to the RR fields 
$B_2^i$, $i=1,2,3$.
\begin{eqnarray}
B_2^1 \wedge \left( \frac{- 2 \epsilon {\tilde \epsilon} \beta^1 }{\beta^2 }
\right)F^b, \ \ 
B_2^2 \wedge \left(\frac{\epsilon \beta^2}{\beta^1}  
\right)(9F^a + 2   F^d+  F^e),&\nonumber\\
B_2^3  \wedge \left( \frac{3 {\tilde \epsilon} n_a^2}{2\beta^1} F^a +     
\frac{n_b^1}{\beta^2}F^b  + \frac{n_c^1}{\beta^2} F^c -
\frac{{\tilde \epsilon} n_d^2}{2\beta^1} F^d
-\frac{{\tilde \epsilon} n_e^2}{2 \beta^1}F^e \right).&
\label{rr1}
\end{eqnarray}
In an orthogonal basis, the rest of the U(1)'s, are the SM hypercharge
\begin{equation}
(3 n_a^2 + 3n_d^2 + 3 n_e^2) \neq 0,\;\ Q^l = n_c^1 (Q_a -3 Q_d -3 Q_e)
-\frac{3 {\tilde \epsilon}\beta^2 ( n_a^2 + n_d^2 + n_e^2)}{2 \beta^1} Q_c  
\label{hyper}
\end{equation}
only when (\ref{hyper}) satisfies the condition, 
\begin{equation}
n_c^1 = \ \frac{{\tilde \epsilon} \beta^2}{2\beta^1}( n_a^2 +\  
n_d^2 +\  n_e^2),
\label{mashyper}
\end{equation}
as well the 
\begin{equation}
U(1)^{(5)} = (-\frac{3}{29}  + \frac{3}{28}) F^a -\frac{1}{29}F^d + 
 \frac{1}{28} F^3 \ .
\label{extrayou1}
\end{equation}
The latter $U(1)$ may be broken by demanding that the open string sector
$ac$ respects $N = 1$ SUSY, giving us a constraint on the tadpole parameter
$n_e^2 = 0$ as well allowing the presence
of $s\nu_R$. Also the presence of (\ref{extrayou1}) gives the 
constraint $n_a^2 = (-28/9)n_d^2$.
In a similar way we can treat the construction of higher order deformation,
with only SM at low energy of order $M_Z$, 
involving the six-stack \cite{mo4} structure of table (\ref{spectrum88}) 
as well
the Pati-Salam GUTS \cite{mo2} of table (\ref{spectrum8001}).

\begin{table}[th]
\begin{center}
{\footnotesize
\begin{tabular}{|c||c|c|c|}
\hline
$N_i$ & $(n_i^1, m_i^1)$ & $(n_i^2, m_i^2)$ & $(n_i^3, m_i^3)$\\
\hline
 $N_a=3$ & $(1/\beta^1, 0)$  &
$(n_a^2,  \epsilon \beta^2)$ & $(3, {\tilde \epsilon}/2)$  \\
\hline
$N_b=2$  & $(n_b^1, -\epsilon \beta^1)$ & $(1/\beta^2, 0)$ &
$({\tilde \epsilon}, 1/2)$ \\
\hline
$N_c=1$ & $(n_c^1, \epsilon \beta^1)$ &
$(1/\beta^2, 0)$  & $(0, 1)$ \\    
\hline
$N_d=1$ & $(1/\beta^1, 0)$ &  $(n_d^2,  2 \epsilon \beta^2)$  
  & $(1, -{\tilde \epsilon}/2)$  \\\hline
$N_e = 1$ & $(1/\beta^1, 0)$ &  $(n_e^2,   \epsilon \beta^2)$  
  & $(1, -{\tilde \epsilon}/2)$  \\
\hline
\end{tabular}
\label{spectrum10}}
\end{center}
\end{table}

Table 1.a. Tadpole solutions of 
D6-branes wrapping numbers. 
The solutions depend 
on five integer parameters, 
$n_a^2$, $n_d^2$, $n_e^2$, $n_b^1$, $n_c^1$,
the NS-background $\beta^i$ and
the phase parameters $\epsilon = \pm 1$, ${\tilde \epsilon} = \pm 1$.

\begin{table}[th] 
\tbl{Low energy fermionic spectrum of the six stack 
string scale 
$SU(3)_C \otimes
SU(2)_L \otimes U(1)_a \otimes U(1)_b \otimes U(1)_c 
\otimes U(1)_d \otimes U(1)_e \otimes U(1)_f$, 
type I 
D6-brane model 
together with its
$U(1)$ charges. Note that at low energies only the SM gauge group 
$SU(3) \otimes SU(2)_L \otimes U(1)_Y$ survives.
\vspace*{1pt}}
{\footnotesize
\begin{tabular}{|c|c|c|c|c|c|c|c|c|c|}
\hline
Matter Fields & & Intersection & $Q_a$ & $Q_b$ & $Q_c$ & $Q_d$ & $Q_e$&
$Q_f$ & Y
\\\hline
 $Q_L$ &  $(3, 2)$ & $I_{ab}=1$ & $1$ & $-1$ & $0$ & $0$ & $0$&
 $0$  & $1/6$ \\\hline
 $q_L$  &  $2(3, 2)$ & $I_{a b^{\ast}}=2$ &  
$1$ & $1$ & $0$ & $0$  & $0$ &$0$ & $1/6$  \\\hline
 $U_R$ & $3({\bar 3}, 1)$ & $I_{ac} = -3$ & 
$-1$ & $0$ & $1$ & $0$ & $0$ & $0$ & $-2/3$ \\\hline    
 $D_R$ &   $3({\bar 3}, 1)$  &  $I_{a c^{\ast}} = -3$ &  
$-1$ & $0$ & $-1$ & $0$ & $0$ &$0$ & $1/3$ \\\hline    
$L^1$ &   $(1, 2)$  &  $I_{bd} = -1$ &  
$0$ & $-1$ & $0$ & $1$ & $0$ &$0$ & $-1/2$  \\\hline    
$L^2$ &   $(1, 2)$  &  $I_{b e} = -1$ &  
$0$ & $-1$ & $0$ & $0$ & $1$ & $0$ & $-1/2$  \\\hline
$L^3$ &   $(1, 2)$  &  $I_{b f} = -1$ &  
$0$ & $-1$ & $0$ & $0$ & $0$ & $1$ & $-1/2$  \\\hline
$N_R^1$ &   $(1, 1)$  &  $I_{cd} = 1$ &  
$0$ & $0$ & $1$ & $-1$ & $0$ & $0$ & $0$  \\\hline    
$E_R^1$ &   $(1, 1)$  &  $I_{c d^{\ast}} = -1$ &  
$0$ & $0$ & $-1$ & $-1$ & $0$ & $0$ &  $1$   \\\hline
  $N_R^2$ &   $(1, 1)$  &  $I_{c e} = 1$ &  
$0$ & $0$ & $1$ & $0$ & $-1$ & $0$ &  $0$ \\\hline
$E_R^{2}$ &   $(1, 1)$  &  $I_{c e^{\ast}} = -1$ &  
$0$ & $0$ & $-1$ & $0$ & $-1$ & $0$ &  $1$   \\\hline
$N_R^3$ &   $(1, 1)$  &  $I_{c f} = 1$ &  
$0$ & $0$ & $1$ & $0$ & $0$ & $-1$ &  $0$ \\\hline
$E_R^3$ &   $(1, 1)$  &  $I_{c f^{\ast}} = -1$ &  
$0$ & $0$ & $-1$ & $0$ & $0$  &$-1$ & $1$ \\\hline
\hline
\end{tabular}
\label{spectrum88}}
\end{table}
\begin{table}[th]
\tbl{
Fermionic spectrum of the $SU(4)_C \times
SU(2)_L \times SU(2)_R$, PS-A class of models together with $U(1)$ charges.
\vspace*{1pt}}
{\footnotesize
\begin{tabular}{|c|c||c|c||c||c|c|}
\hline
Fields &Intersection  & $\bullet$ $SU(4)_C \times SU(2)_L \times SU(2)_R$
$\bullet$&
$Q_a$ & $Q_b$ & $Q_c$ & $Q_d$ \\[1ex]
\hline
 $F_L$& $I_{ab^{\ast}}=3$ &
$3 \times (4,  2, 1)$ & $1$ & $1$ & $0$ &$0$ \\
 ${\bar F}_R$  &$I_{a c}=-3 $ & $3 \times ({\bar 4}, 1, 2)$ &
$-1$ & $0$ & $1$ & $0$\\
 $\chi_L$& $I_{bd} = -12$ &  $12 \times (1, {\bar 2}, 1)$ &
$0$ & $-1$ & $0$ & $1$ \\    
 $\chi_R$& $I_{cd} = -12$ &  $12 \times (1, 1, {\bar 2})$ &
$0$ & $0$ & $-1$ & $-1$ \\\hline
 $\omega_L$ & $I_{aa^{\ast}}$ &  $12 \beta^2
 {\tilde \epsilon }
 \times (6, 1, 1)$ & $2{\tilde \epsilon }$ & $0$ & $0$ & $0$ \\
 $z_R$& $I_{aa^{\star}}$ & $6  \beta^2 {\tilde \epsilon } \times 
({\bar 10}, 1, 1)$ &
$-2{\tilde \epsilon }$ & $0$ & $0$ &$0$  \\
 $s_L$& $I_{aa^{\ast}}$ & $24  \beta^2 {\tilde \epsilon }
 \times (1, 1, 1)$ &
$0$ & $0$ & $0$ &$-2{\tilde \epsilon }$  \\
\hline
\end{tabular}
\label{spectrum8001} }
\end{table}
\section*{Acknowledgments}
I am grateful to the organizers of SP2002 
for providing me with
financial support.

\end{document}